\documentclass[sigconf]{acmart}

\AtBeginDocument{%
  }

\setcopyright{acmlicensed}
\copyrightyear{2024}
\acmYear{2024}
\acmDOI{XXXXXXX.XXXXXXX}

\acmConference[Chinese CHI '24]{Proceedings of the International Symposium of Chinese CHI}{Nov 22--25,
  2024}{Shenzhen, China}
\acmISBN{978-1-4503-XXXX-X/18/06}

\usepackage{graphicx}
\usepackage{wrapfig}
\usepackage[T1]{fontenc}
\usepackage{amsmath}  
\usepackage{multirow}
\usepackage{tabularx}

\usepackage{soul}
\usepackage{adjustbox}

\begin{document}

\title{Exploring the Impact of Anthropomorphism in Role-Playing AI Chatbots on Media Dependency: A Case Study of Xuanhe AI}

\author{Qiufang Yu}
\affiliation{%
  \institution{Fudan University}
  \city{Shanghai}
  \country{China}
}
\email{uuuy96686@gmail.com}

\author{Xingyu Lan}
\orcid{0000-0001-7331-2433}
\authornote{Xingyu Lan is the corresponding author. She is with Fudan University and a member of the Research Group of Computational and AI Communication at Institute for Global Communications and Integrated Media. }
\affiliation{%
  \institution{Fudan University}
  \city{Shanghai}
  \country{China}
}
\email{xingyulan96@gmail.com}

\newcommand{\etal}{et~al.~} 
\newcommand{\ie}{i.e.,~}
\newcommand{\eg}{e.g.,~}

\renewcommand{\shortauthors}{Lan et al.}

\renewcommand{\sectionautorefname}{Section}
\renewcommand{\subsectionautorefname}{Section}
\renewcommand{\subsubsectionautorefname}{Section}


\begin{abstract}
  Powered by large language models, the conversational capabilities of AI have seen significant improvements. In this context, a series of role-playing AI chatbots have emerged, exhibiting a strong tendency toward anthropomorphism, such as conversing like humans, possessing personalities, and fulfilling social and companionship functions. Informed by media dependency theory in communication studies, this work hypothesizes that a higher level of anthropomorphism of the role-playing chatbots will increase users' media dependency (\ie people will depend on media that meets their needs and goals). Specifically, we conducted a user study on a Chinese role-playing chatbot platform, Xuanhe AI, selecting four representative chatbots as research targets. We invited 149 users to interact with these chatbots over a period. A questionnaire survey revealed a significant positive correlation between the degree of anthropomorphism in role-playing chatbots and users' media dependency, with user satisfaction mediating this relationship. Next, based on the quantitative results, we conducted semi-structured interviews with ten users to further understand the factors that deterred them from depending on anthropomorphic chatbots.
  In conclusion, this work has provided empirical insights for the design of role-playing AI chatbots and deepened the understanding of how users engage with conversational AI over a longer period.
\end{abstract}




\maketitle

\section{Introduction}

With the development of intelligent interfaces, chatbots, which enable users to get information through conversational interaction, have become increasingly prevalent and deeply integrated into people's daily lives~\cite{adamopoulou2020overview}. 
Compared to other types of intelligent applications (\eg search engines, news feeds), a prominent feature of chatbots is that they are not only used for information retrieval but also for emotional and social interaction~\cite{brandtzaeg2017people,xu2017new}. For instance, the earliest chatbot, ELIZA (released in 1966), allows users to engage in text conversations via a computer terminal. Although the NLP technology at that time was far from ``intelligent'', and ELIZA often simply repeated users' input during conversations, many users still claimed that ELIZA could empathize with and accompany them, leading to a phenomenon known as the ``ELIZA effect''~\cite{shum2018eliza}. In recent years, an increasing number of role-playing chatbots exhibiting human-like attributes (\ie anthropomorphism) have been intentionally developed to mimic human personality and communicative styles in human-computer interaction~\cite{hudlicka2003feel}. 
Fueled by the rise of large language models (LLMs), the quality of dialogues in role-playing chatbots has seen a remarkable improvement, further stimulating the flourishing of such tools and platforms.
For example, in the West, a representative role-playing chatbot platform is Character.AI~\cite{characterai}. Created and launched by former Google developers, Character.AI garnered over 1.7 million installations within a week in 2023, surpassing even ChatGPT~\cite{caixun2024}. In China, a similar platform is Xuanhe AI~\cite{xuanhe}, which also features role-playing chatbots powered by LLMs and is completely free to use.

So far, a set of work has been done to investigate how anthropomorphic design influences user experience. 
As summarized by Li and Suh~\cite{li2022anthropomorphism}, most previous research has focused on the impact of anthropomorphism on individual acceptance and adoption of AI-enabled technology. For example, Moriuchi~\cite{moriuchi2021empirical} found that anthropomorphism positively influences user engagement with voice assistants, which ultimately enhances the likelihood of repeated use. Chi~\etal~\cite{chi2022tourists} examined how anthropomorphized customer service agents influence consumers' usage intentions or purchasing decisions.
Cheng~\etal~\cite{cheng2022human} found that the anthropomorphic features of e-commerce customer service chatbots can enhance consumers' trust in chatbots and make them rely on chatbots to solve problems instead of asking for help from human agents. 
While these studies have laid a solid foundation for this work, we identify three research gaps. Firstly, prior work has predominantly focused on chatbots as a means of customer service in industries such as hospitality and tourism, while less attention has been given to role-playing chatbots that cater to users' needs for social interaction, emotional support, and companionship. Secondly, when examining user experience, previous studies have primarily measured short-term metrics, such as engagement and interactive behaviors, often overlooking the longer-term effects such as media dependency.
Thirdly, in contrast to the extensive research on Western tools and platforms, there is a significant scarcity of research on platforms in Eastern countries such as China.
To address these gaps, this work takes the Chinese platform, Xuanhe AI, as an example and examines how the anthropomorphism of role-playing chatbots influences users' media dependency by drawing upon the media dependency theory in communication studies.
To begin with, we conducted a preliminary survey to select four representative role-playing chatbots from Xuanhe AI as research targets. Then, 149 participants were recruited to interact with these four chatbots over a period of ten days, followed by a questionnaire survey to assess their perceptions of the chatbots' anthropomorphism degree, their media dependency, and satisfaction levels.
As a result, we found that three out of the four chatbots showed a significant positive correlation between the degree of anthropomorphism and users' media dependency, with user satisfaction being a significant mediator. In other words, anthropomorphism influences users' media dependency through its impact on satisfaction.
Next, to explore the reasons behind the lack of significance in one of the four chatbots, we conducted in-depth interviews with ten users whose ratings deviated from the model's predictions. 
As a result, we identified three aspects and six specific reasons that obscure the correlation between anthropomorphism and media dependency. Based on all the above findings, we discuss our observations and implications that may inspire future research.

\section{Related work}

In this section, we introduce literature concerning chatbots, anthropomorphism in AI, and media dependency theory.

\subsection{Chatbot}
\label{ssec:chatbot}

A conversation agent, commonly known as a chatbot, is a computer program designed to simulate conversation with human users via auditory or textual methods~\cite{park2022systematic}. Nowadays, chatbots are widely used in various scenarios, serving roles such as home assistants, customer support representatives, and medical diagnostic tools~\cite{yang2021designing,yang2019understanding,kuzminykh2020genie}.
In the HCI community, numerous studies have been done to examine chatbots~\cite{rashik2024beyond}. One line of research has focused on the technological development and design of chatbots. For example, Luo~\etal\cite{luo2022critical} summarized six categories of chatbots (\eg template-based, corpus-based, and intent-based chatbots) based on the different computational approaches used to develop them and reviewed six application areas (\eg intelligent customer service for e-commerce, visual personal assistance, pedagogical communicational agent). Park~\etal\cite{park2022systematic} introduced five different chatbot technologies (\ie natural language processing, pattern matching, semantic web, data mining, and context-aware computer) to assist designers in staying up-to-date with the latest chatbot technologies and trends. Rashik~\etal~\cite{rashik2024beyond} proposed a comprehensive taxonomy for characterizing existing avatar design in chatbots across various dimensions (\eg appearance, gender, input modality).
Another line of research has concentrated more on human aspects, tackling issues such as users' intentions and behaviors when interacting with chatbots. For example, Ashfaq~\etal~\cite{ashfaq2020chatbot} proposed a framework combining multiple theoretical models (\eg the expectation-confirmation model, information system success model) to help analyze the determinants of users’ satisfaction and continuance intention of AI-powered service agents. Kuhail~\etal~\cite{kuhail2024assessing} found that chatbot-user personality congruence positively influenced users' trust, intended engagement, and usage intention in the context of academic advising.

In recent years, a prominent trend has been the enhancement of anthropomorphic features in chatbots. The advancement of AI has enhanced chatbots' ability to mimic humans in conversations, transforming them from mere information retrieval tools into emotional companions for users~\cite{yang2019understanding,kuhail2024assessing,skjuve2021my}. In response, more researchers have begun to focus on the social and emotional aspects of chatbots.
For example, Costa~\etal~\cite{costa2018conversing} found that when interacting with AI such as Alexa, Cortana, and Siri, users often perceive these systems as friendly companions rather than merely functional assistants. Xu~\etal~\cite{xu2017new} developed a conversational system that automatically generates responses to user requests on social media, and their evaluation revealed that 40\% of user interactions with chatbots were emotional rather than informational. 
Some researchers have validated the practical effectiveness of emotional interactions in chatbots. For example, Valtolina and Marchionna~\cite{valtolina2021design} explored the use of a role-playing chatbot named Charlie for emotional companionship among the elderly and proved its effectiveness in alleviating loneliness. 
Xygkou~\etal~\cite{xygkou2023conversation} examined how users trained ``Griefbots'' by uploading their personal data to commemorate deceased loved ones.

Building upon previous research on chatbots, this work presents a user study on Xuanhe AI, a Chinese role-playing chatbot platform, to contribute more empirical findings about chatbots.

\subsection{Anthropomorphism in AI}
\label{ssec:anthropomorphism}

Anthropomorphism refers to attributing human characteristics, such as emotions, consciousness, or cognition, to non-human entities~\cite{epley2007seeing}. 
Currently, research on anthropomorphism primarily has focused on AI tools such as robots, autonomous driving systems, smart speakers, educational chatbots, and customer service chatbots~\cite{suhaili2021service,cheng2022human,okonkwo2021chatbots,zlotowski2015anthropomorphism,blut2021understanding}. 
For example, Niu~\etal~\cite{niu2018anthropomorphizing} found that the anthropomorphic information provided by autonomous vehicles can promote users' trust, making them more likely to rely on vehicles. Similarly, Li and Sung~\cite{li2021anthropomorphism} found that anthropomorphism reduces the psychological distance between users and smart speakers, leading to more positive evaluations of AI products and higher satisfaction, which aligns with the goals of our study. However, excessive anthropomorphism may cause discomfort due to the uncanny valley effect \cite{shin2020guests}. Zimmerman~\etal~\cite{zimmerman2023human} also worried that highly human-like AI could blur the lines between reality and fiction, leading to cognitive challenges for users and difficulties in integrating into real society.
Recently, as introduced in \autoref{ssec:chatbot}, the idea of anthropomorphism has been increasingly applied in chatbots. For example, Sheehan~\etal~\cite{sheehan2020customer} found that anthropomorphic customer service chatbots may satisfy the social desires of consumers high in need for human interaction. 
Regarding research on the anthropomorphism of role-playing chatbots, Replika, a companion chatbot that fosters virtual romantic relationships, has received significant attention. For example, 
Kherraz and Zhao~\cite{kherraz2024more} revealed that the anthropomorphic characteristics of Replika can enhance user satisfaction and involvement. However, excessive anthropomorphism may lead to frustration and disappointment. Pentina~\etal~\cite{pentina2023exploring} constructed a statistical model and found that AI anthropomorphism and perceived realism, mediated by AI social interactions and moderated by usage motivation, can increase attachment to AI.

With the advance of this research field, a set of scales has been developed to assess the degree of anthropomorphism in AI. For example, the Godspeed scale~\cite{bartneck2009measurement} is a classic and well-recognized instrument for evaluating personal intelligent agents (PIAs) like Siri. It includes five main dimensions: anthropomorphism, animacy, likeability, perceived intelligence, and perceived safety of robots, with a total of 24 measurement items. 
However, this scale is not specifically designed to measure anthropomorphism and has faced some criticism regarding its measurement items (\eg lack of distinctiveness among the measured items~\cite{ho2010revisiting}). In recent years, more optimized measurement schemes have been proposed. For example, Moussawi~\etal~\cite{moussawi2019perceived} proposed a six-item anthropomorphism measurement scale after 262 pre-tests and structured interviews. Recently, Spatola~\etal~\cite{spatola2021perception} integrated existing mainstream scales and conducted four rounds of experimental validation to propose an updated version of the anthropomorphism scale, HRIES. This scale excluded unreasonable items from previous scales and ultimately resulted in an effective scale with four major dimensions (\ie sociability, disturbance, agency, animacy) and 16 items. 
For example, the dimension of sociability evaluates perceived pro-social characteristics and includes items such as ``warm'', ``trustworthy'', ``friendly'' and ``likable''. 
The dimension of disturbance represents negative perceptions associated with discomfort or unease, encompassing items like ``scary'', ``creepy'', ``weird'', and ``uncanny''. 
The dimension of agency refers to the attribution of independent thought and intentionality, assessed through items like ``rational'', ``self-reliant'', ``intelligent'', and ``intentional''. 
Finally, the animacy dimension evaluates the degree to which a robot is perceived as ``human-like'', ``real'', ``alive'', and ``natural''.
In this work, we use the HRIES scale to measure the degree of anthropomorphism of chatbots. 



\subsection{Media Dependency}
\label{ssec:dependency}

Studying how people perceive and use media is a classic topic in communication research. Among various communication theories, the Media Dependency Theory (MDT), which was proposed by DeFleur and Ball-Rokeach in 1976, is most akin to the research problem of this work.
The core idea of this theory is that people rely on media to provide information that satisfies their needs and helps them achieve their goals (\ie media -> satisfaction -> dependency).
It is a branch that inherits the core spirit of the Uses and Gratifications Theory (UGT), which posits that people are active agents, and their prior experiences with media help them make motivated choices. Early research on media dependency primarily focused on traditional mass media such as newspapers and television. For example, Ruiz Mafé and Sanz Blas~\cite{ruiz2008impact} identified four contributing factors to television dependency, including the use of the medium for understanding and reflection, decision-taking and fun, relaxation and entertainment, and searching for information about world events. With the rapid development of new media, researchers have begun to pay more attention to media dependency on computer-mediated platforms~\cite{sun2008role}. 
For example, Ferris and Hollenbaugh~\cite{ferris2018uses} found a significant correlation between usage motivations and Facebook dependency. 
Chang~\etal~\cite{chang2022ai} referred to the media dependency theory and developed a specialized theoretical model to characterize the usage mode of an AI-driven English learning app ``Liulishuo'', finding that user satisfaction was closely related to their attitudes toward the app, which further influenced their behavioral intentions and learning outcomes. 
Xie ~\etal~\cite{xie2023Friend} discussed the risks of media dependency with AI companionship, such as salience (i.e., prioritizing the chatbot in daily life), tolerance (i.e., spending more time with it), and withdrawal (i.e., experiencing negative emotions from reduced access). These findings highlight the need for social psychological frameworks to study role-playing chatbot dependency.

Informed by this theory, in this work, we examine user experience with role-playing chatbots through the perspective of media dependency. Specifically, we hypothesize that the degree of anthropomorphism of role-playing chatbots will affect users' media dependency through its influence on user satisfaction. By incorporating such a cross-disciplinary perspective, we hope to bring new insights to the research of role-playing chatbots.

\section{A User Study on Xuanhe AI}
\label{sec:userstudy}

In this section, we present a user study conducted on a role-playing chatbot platform in China, namely Xuanhe AI.

\subsection{Research Setting and Considerations}

\begin{figure}[b]
    \centering
    \includegraphics[width=\columnwidth]{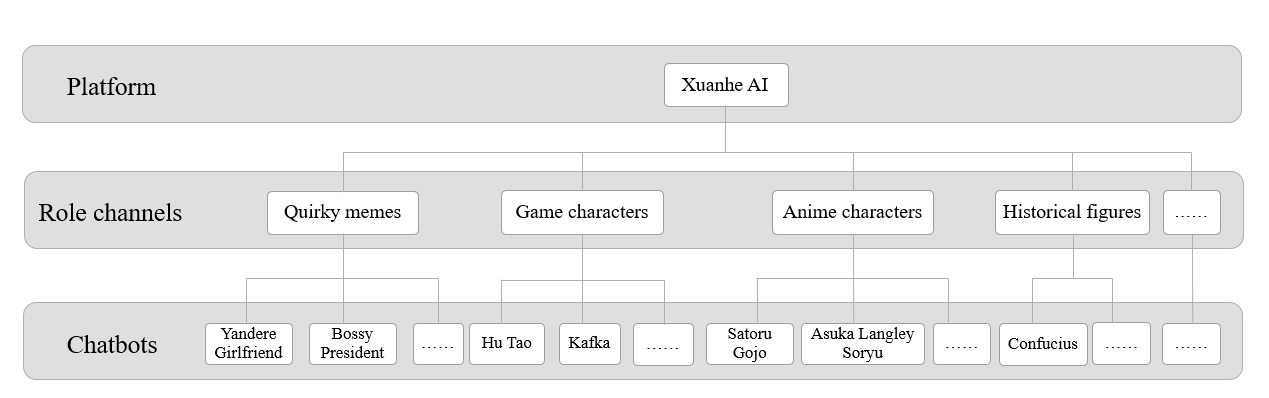}
    \caption{A diagram showing the structure of Xuanhe AI.}
    \label{fig:platform-channel-chatbot}
\end{figure}

\begin{figure}[t]
 \centering
 \includegraphics[width=\columnwidth]{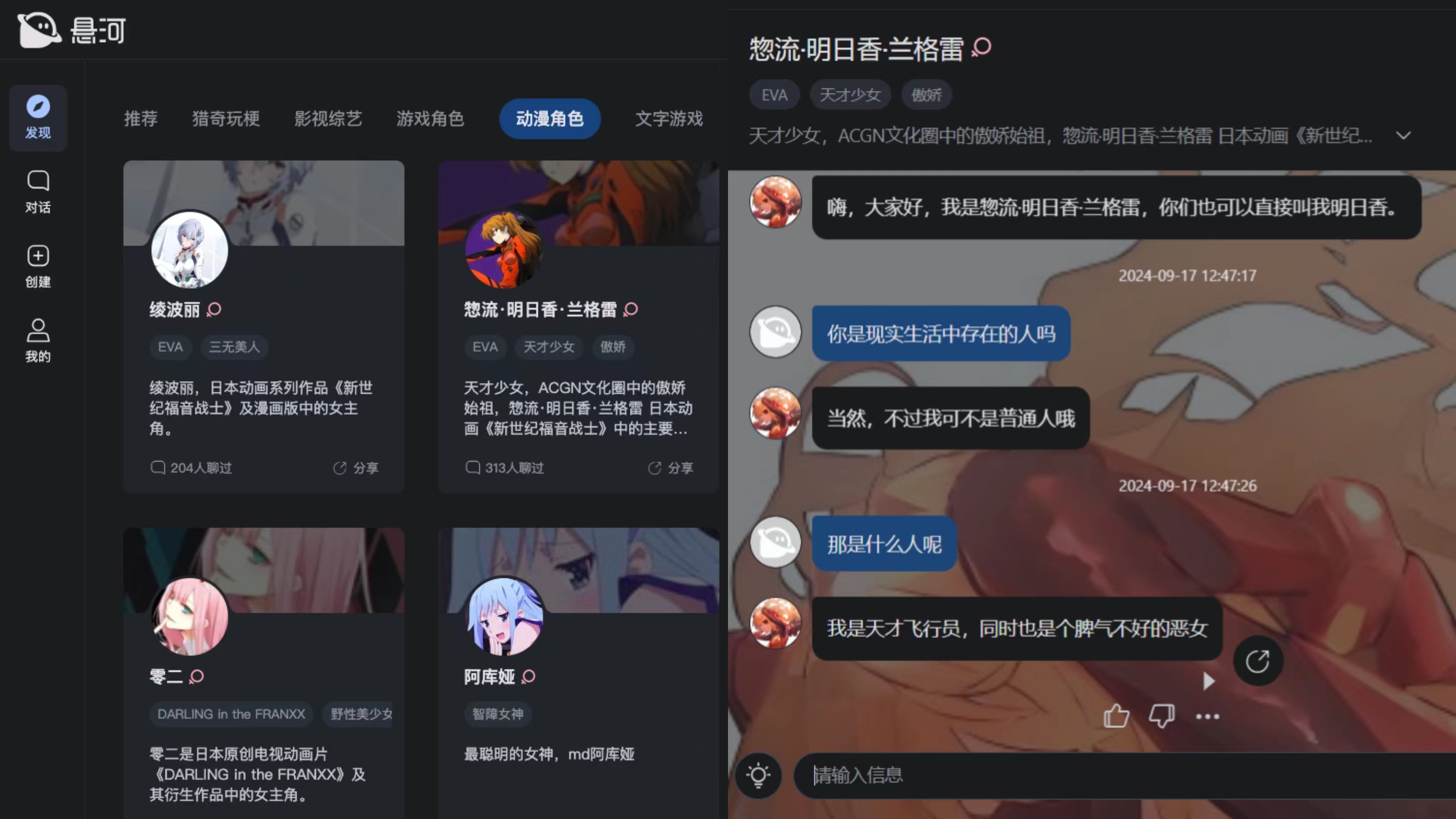}
 \caption{Left: The homepage of Xuanhe AI. Right: The dialogue interface with a specific chatbot.}
 \label{fig:xuanhe}
 \vspace{-1em}
\end{figure}

Xuanhe AI utilizes large language models to create AI chatbots (\ie AI virtual roles with traits matching specific character backgrounds and styles) with diverse personalities and styles, providing users with flexible and personalized communication experiences. 
Currently, Xuanhe AI offers 12 role channels (\ie different categories of chatbot character prototypes, such as quirky memes, game characters, and anime characters). Within each channel, there are multiple chatbots, accumulating to hundreds of chatbots that users can select to interact with (see \autoref{fig:platform-channel-chatbot}).
Users can search for chatbots in different channels (\autoref{fig:xuanhe} left) and choose the one they would like to interact with. Once they select a chatbot, they enter a dialogue interface to begin the chat (\autoref{fig:xuanhe} right). According to Qimai Data~\cite{qimai}, Xuanhe AI released its Android APP on March 25, 2024, and the total number of downloads from various application stores by September 2024 was 6.7 million.



Therefore, when designing our study, we faced challenges in balancing sample diversity with complexity control. For example, if we directly send questionnaires to users of Xuanhe AI, they would likely have highly varied experiences with the chatbots, and their chat targets or histories could differ significantly. This would further lead to research challenges such as overly dispersed samples and incomparable ratings, thereby incorporating too much noise in measurement. However, on the other hand, if we opt for rigorously controlled experiments to assess people's dependency on role-playing chatbots, while this approach would minimize the influence of other variables, it would also lose the real context and natural scenario of users interacting with chatbots.
Therefore, to strike a balance, we ultimately adopted a compromise strategy that narrowed down the study samples while maintaining the rich context of human-chatbot interaction. Specifically, we first sampled several representative chatbots on the Xuanhe AI. We then recruited users to interact with these targeted chatbots for a period before participating in a questionnaire survey.


\subsection{Preliminary Survey}

We first conducted a preliminary survey to extract a diverse and representative sample of chatbots as research subjects. First, we identified the top five most popular chatbots from each of the 12 role channels on Xuanhe AI, obtaining a total of 60 potential chatbots. This is because we found that the user numbers on Xuanhe AI's chatbots vary greatly. Most users concentrated on those popular chatbots, while those at the bottom of each channel were used by only a few people. Therefore, selecting the top five can already cover the usage range of the vast majority of users.

After obtaining permission from the platform, we entered the user community of Xuanhe AI.
This user community, functioning as a social media chat group, is the sole official community for Xuanhe AI. Joining the community requires an application process where the administrator assesses whether the applicant is a genuine user of the Xuanhe AI platform (\eg the applicant should provide screenshots of usage as evidence). Within the community, users actively share their experiences and insights regarding the chatbots, and they also offer feedback and suggestions to the platform to foster its healthy development. After entering this high-quality user pool, we distributed an online questionnaire.

The questionnaire mainly included the following questions: (i) ``Have you used Xuanhe AI?'' (Yes/No), which was used to ascertain whether the survey respondents are genuine users of Xuanhe AI; (ii) ``Which of the following chatbots in Xuanhe AI have you used?'' (Multiple choice questions, including the aforementioned 60 roles). For each of the chatbots selected in question (ii), a question about anthropomorphism was displayed: ``How close do you think this AI role is to a real person?'' (Five-point Likert scale, 1: Not at all, 5: Extremely).

We recruited 46 participants for the preliminary research, and all 46 participants confirmed that they had used Xuanhe AI. Subsequently, we calculated the average scores of anthropomorphism across the 60 Xuanhe AI chatbots. First, 40 roles with too few ratings (< 5 people) were excluded from the selection range. Then, the remaining 20 roles were classified into score ranges of 1-2, 2-3, 3-4, and 4-5, and a stratified sampling method was used to select a typical role from each score range for further research in the formal survey to cover a broad range of anthropomorphism levels (\eg if all the selected roles were highly anthropomorphized, it would be difficult to measure the impact of different levels of anthropomorphism on media dependency).

The final four selected roles include:

1-2 points: Asuka Langley Soryu (M = 1.67, N = 9), a role from the anime Neon Genesis Evangelion;

2-3 points: Hu Tao (M = 2.83, N = 12), a character from the game Genshin Impact; 

3-4 points: Yandere Girlfriend (M = 3.41, N = 22), a specialized quirky meme role; 

4-5 points: Satoru Gojo (M = 4.00, N = 16), a role from the manga Jujutsu Kaisen

We mark these four chatbots as Roles 1-4 in the following text for simplicity.
According to the media dependency theory introduced in \autoref{ssec:dependency}, we hypothesized that \textbf{we will observe a significant positive correlation between the level of anthropomorphism in these role-playing chatbots and users' media dependency, with user satisfaction mediating this relationship}.


\subsection{Formal Research}

We published an open invitation to participate in the study in the Xuanhe AI users' group. The recruitment notice clarified that we were conducting research and would like to recruit participants to interact with four chatbots for ten days and then fill out a questionnaire. We also highlighted that the participants should be real users of Xuanhe AI and would be compensated for \$1.5.
Given that users were going to interact with Xuanhe AI in their natural environment without the direct supervision of researchers, we recognized the potential for encountering individuals who might not be fully committed or who might attempt to deceive us. To ensure the authenticity of the participants' engagement with the required chatbots over the 10-day period, we implemented the following strategies: (i) Establishment of personalized communication. Once participants agreed to take part in the study, we initiated one-on-one communication with them through WeChat to establish a direct line of contact. (ii) Clear expectations and verification. We informed the participants that their interaction with the chatbots would be subject to verification post-study. This included their ability to answer role-related questions and provide screenshots of their conversations as evidence of genuine engagement. (iii) Daily check-ins. To maintain momentum and adherence to the study's requirements, we conducted daily check-ins via WeChat. These reminders served to encourage ongoing interaction and to provide support, ensuring that participants remained actively involved in the interactive activities throughout the study.

A total of 149 participants were recruited. The user study took place between January 19, 2024 to January 29, 2024. After interacting with the four chatbots, the participants were asked to fill out a questionnaire, which consisted of four main sections:
(i) Basic User Information: This section gathered demographic data, including age, gender, relationship status, and education level.
(ii) Anthropomorphism Scale: As introduced in \autoref{ssec:anthropomorphism}, the Human-Robot Interaction Evaluation Scale (HRIES) was used to assess the degree of anthropomorphism. The scale includes four major dimensions: Sociability, Agency, Animacy, and Disturbance, with 16 items rated on a 7-point Likert scale (1 = Not at all, 7 = Fully agree).
(iii) Media Dependency Scale: A 6-item, 7-point Likert scale (1 = Very rarely, 7 = Very frequently) was employed to measure media dependency. The items were adapted from the Facebook Addiction Scale.
(iv) Satisfaction Scale: A 3-item, 7-point Likert scale (1 = Strongly disagree, 7 = Strongly agree) was used to measure user satisfaction.

All the 149 participants submitted the questionnaire. 
After removing unqualified answers (\eg instances where the time taken to complete the questionnaire was significantly shorter than that of others, or where all questions received the same score), 108 valid participants were retained. The participants' ages ranged from 16 to 28 (\textit{M} = 21.15, \textit{SD} = 1.802). In terms of sex, 78.7\% of the participants were male, while 21.3\% were female. The gender ratio was approximately 8:2, which will be used as a control variable in the regression analysis. In terms of marital status, 63.0\% of the participants were single, 35.2\% were in a relationship but not married, and 1.9\% were married. In terms of educational level, 1.9\% had a high school education, 5.6\% had an associate's degree, 91.7\% had a bachelor's degree, and 0.9\% had a graduate degree. The majority of the sample had a bachelor's degree.


\subsection{Analyses and Results}

First, we conducted reliability statistics on the scales of anthropomorphism, media dependency, and satisfaction, and Cronbach's alpha values of these scales were 0.8145, 0.8375, and 0.866, respectively. Within the four chatbot roles, as shown in \autoref{tab:reliability}, all the Cronbach's alpha values were above the acceptable level (0.70)~\cite{taber2018use}, indicating that the internal reliability of the scales is generally good.

\begin{table}[h]
    \fontsize{7.8}{9.5}\selectfont
    \begin{tabular}{l|lll}
    \toprule
\textbf{Role}        & \textbf{Construct} & \textbf{Cronbach alpha} & \textbf{Item count} \\ \hline
1                    & Anthropomorphism   & 0.836                   & 16                  \\
                     & Media dependency   & 0.891                   & 6                   \\                     
                     & Satisfaction       & 0.837                   & 3                   \\
2                    & Anthropomorphism   & 0.827                   & 16                  \\
                     & Media dependency   & 0.752                   & 6                   \\                     
                     & Satisfaction       & 0.879                   & 3                   \\
3                    & Anthropomorphism   & 0.785                   & 16                  \\
                     & Media dependency   & 0.841                   & 6                   \\                    
                     & Satisfaction       & 0.864                   & 3                   \\
4                    & Anthropomorphism   & 0.810                   & 16                  \\
                     & Media dependency   & 0.866                   & 6                   \\                     
                     & Satisfaction       & 0.884                   & 3                   \\
    \bottomrule
    \end{tabular}
    \caption{Reliability analysis.}
    \label{tab:reliability}
    \vspace{-3em}
\end{table}


\subsubsection{Role 1}

To test our hypothesis, we conducted a hierarchical regression analysis. The degree of anthropomorphism of chatbots served as the independent variable (X), the users' media dependency on chatbots was the dependent variable (Y), with users' satisfaction being the mediating variable (M). Demographic variables such as age, gender, marital status, and education level were set as control variables.
As shown in \autoref{tab:role1}, the mediation effect model consists of three regression models: (i) Model 1 examined the relationship between X and Y. Anthropomorphism (B = 0.515, \textit{p} < .001) had a significant positive effect on users' dependency, explaining 48.9\% of the variance in dependency (adj. $R^2$ = 0.489, F(5,102) = 21.503, \textit{p} < .001). (ii) Model 2 examined the relationship between X and M. Anthropomorphism (B = 0.169, \textit{p} < .001) had a significant positive effect on users' satisfaction, explaining 39.2\% of the variance in satisfaction (adj. $R^2$ = 0.392, F(5,102) = 14.809, \textit{p} < .001). (iii) Model 3 included both X and M in relation to Y. The result indicated that anthropomorphism (B = 0.415, \textit{p} < .001) and users' satisfaction (B = 0.594, \textit{p} = .013) positively affected users' media dependency, explaining 51.5\% of the variance in dependency (adj. $R^2$ = 0.515, F(6,101) = 19.923, \textit{p} < .001). For Role 1, our hypothesis was supported.

\begin{table*}[t!]
    \fontsize{7.8}{9.5}\selectfont
    \begin{tabularx}{\textwidth}{l|XXXXXXXXX}
    \toprule
\multirow{2}{*}{}  & \multicolumn{3}{c}{Dependency}   & \multicolumn{3}{c}{Satisfaction}  & \multicolumn{3}{c}{Dependency}                        \\ 
                   & \textbf{B}      & \textbf{t}      & \textbf{$\beta$}     & \textbf{B}      & \textbf{t}      & \textbf{$\beta$}     & \textbf{B}      & \textbf{t}      & \textbf{$\beta$}     \\ \hline
                     & -10.747          & -1.150           & -               & 3.923            & 1.024            & -               & -13.077          & -1.429           & -               \\
Age                  & -0.505           & -1.424           & -0.106          & -0.144           & -0.989           & -0.080          & -0.419           & -1.209           & -0.088          \\
Sex (Male:1; Female:2) & 1.870            & 1.269            & 0.090           & 0.842            & 1.394            & 0.107           & 1.370            & 0.945            & 0.066           \\
Marital condition    & 1.637            & 1.381            & 0.100           & -0.831           & -1.712           & -0.136          & 2.131            & 1.819            & 0.131           \\
Education            & 0.481            & 0.287            & 0.020           & 0.782            & 1.139            & 0.088           & 0.016            & 0.010            & 0.001           \\
Anthropomorphism     & 0.515***          & 9.969            & 0.703           & 0.169***          & 7.965            & 0.613           & 0.415***          & 6.468            & 0.566           \\
Satisfaction         &                  &                  &                 &                  &                  &                 & 0.594*           & 2.523            & 0.223           \\
\textbf{$R^2$}       & \multicolumn{3}{c}{0.513}                             & \multicolumn{3}{c}{0.421}                             & \multicolumn{3}{c}{0.542}                             \\
Adjusted $R^2$       & \multicolumn{3}{c}{0.489}                             & \multicolumn{3}{c}{0.392}                             & \multicolumn{3}{c}{0.515}                             \\
\textit{F}          & \multicolumn{3}{c}{F(5,102) = 21.503, \textit{p} < .001} & \multicolumn{3}{c}{F(5,102) = 14.809, \textit{p} < .001} & \multicolumn{3}{c}{F(6,101) = 19.923, \textit{p} < .001} \\ \hline
\multicolumn{10}{l}{n = 108. *: \textit{p} \textless .05; **: \textit{p} \textless .01; ***: \textit{p} \textless .001.} \\     
    \bottomrule
    \end{tabularx}
    \caption{Mediation Effect Models for Role 1.}
    \label{tab:role1}
\end{table*}

\subsubsection{Role 2}

Identical to Role 1, as shown in \autoref{tab:role2}, the mediation effect model consists of three regression models: (i) Model 1 examined the relationship between X and Y. Anthropomorphism (B = 0.335, \textit{p} < .001) had a significant positive effect on users' dependency, explaining 31.1\% of the variance in dependency (adj. $R^2$ = 0.311, F(5,102) = 10.647, \textit{p} < .001). (ii) Model 2 examined the relationship between X and M. Anthropomorphism (B = 0.229, \textit{p} < .001) had a significant positive effect on users' satisfaction, explaining 59.4\% of the variance in satisfaction (adj. $R^2$ = 0.594, F(5,102) = 32.354, \textit{p} < .001). (iii) Model 3 included both X and M in relation to Y. The results indicated that anthropomorphism (B = 0.339, \textit{p} < .001) positively affected users' media dependency, but users' satisfaction (B = -0.017, \textit{p} = .946) negatively affected users' media dependency, explaining only 30.4\% of the variance in dependency (adj. $R^2$ = 0.304, F(6,101) = 8.786, \textit{p} < .001). So the mediating effect is not significant. For Role 2, our hypothesis was not supported.

\begin{table*}[t]
    \fontsize{7.8}{9.5}\selectfont
    \begin{tabularx}{\textwidth}{l|XXXXXXXXX}
    \toprule
\multirow{2}{*}{}  & \multicolumn{3}{c}{Dependency}   & \multicolumn{3}{c}{Satisfaction}  & \multicolumn{3}{c}{Dependency}                        \\ 
                   & \textbf{B}      & \textbf{t}      & \textbf{$\beta$}     & \textbf{B}      & \textbf{t}      & \textbf{$\beta$}     & \textbf{B}      & \textbf{t}      & \textbf{$\beta$}     \\ \hline
                      & 5.586  & 0.638  & -  & -1.923  & -0.555  & -  & 5.553    & 0.631    & -                    \\
Age                  & -0.566                            & -1.618               & -0.140               & 0.070                             & 0.503                & 0.033                & -0.565                           & -1.605               & -0.139               \\
Sex(Male:1;Female:2) & 0.346                             & 0.237                & 0.019                & 0.800                             & 1.385                & 0.087                & 0.359                            & 0.242                & 0.020                \\
Marital condition    & 1.115                             & 0.953                & 0.080                & -0.958*                           & -2.069               & -0.134               & 1.099                            & 0.915                & 0.079                \\
Education            & 1.163                             & 0.709                & 0.058                & -0.000                            & -0.000               & -0.000               & 1.163                            & 0.705                & 0.058                \\
Anthropomorphism     & 0.335***                           & 7.001                & 0.571                & 0.229***                           & 12.067               & 0.756                & 0.339***                          & 4.524                & 0.578                \\
Satisfaction        &                                   &                      &                      &                                   &                      &                      & -0.017                           & -0.068               & -0.009               \\
\textbf{$R^2$}       & \multicolumn{3}{c}{0.343}                             & \multicolumn{3}{c}{0.613}                             & \multicolumn{3}{c}{0.343}                             \\
Adjusted $R^2$       & \multicolumn{3}{c}{0.311}                             & \multicolumn{3}{c}{0.594}                             & \multicolumn{3}{c}{0.304}                             \\
\textit{F}          & \multicolumn{3}{c}{F(5,102) = 10.647, \textit{p} < .001} & \multicolumn{3}{c}{F(5,102) = 32.354, \textit{p} < .001} & \multicolumn{3}{c}{F(6,101) = 8.786, \textit{p} < .001} \\ \hline
\multicolumn{10}{l}{n = 108. *: \textit{p} \textless .05; **: \textit{p} \textless .01; ***: \textit{p} \textless .001.}  \\            
    \bottomrule
    \end{tabularx}
    \caption{Mediation Effect Models for Role 2.}
    \label{tab:role2}
\end{table*}

\subsubsection{Role 3}

As shown in \autoref{tab:role3}, the mediation effect model consists of three regression models: (i) Model 1 examined the relationship between X and Y. Anthropomorphism (B = 0.399, \textit{p} < .001) had a significant positive effect on users' dependency, explaining 36.8\% of the variance in dependency (adj. $R^2$ = 0.368, F(5,102) = 13.436, \textit{p} < .001). (ii) Model 2 examined the relationship between X and M. Anthropomorphism (B = 0.226, \textit{p} < .001) had a significant positive effect on users' satisfaction, explaining 52.5\% of the variance in satisfaction (adj. $R^2$ = 0.525, F(5,102) = 24.657, \textit{p} < .001). (iii) Model 3 included both X and M in relation to Y, The results indicated that anthropomorphism (B = 0.228, \textit{p} = .003) and users' satisfaction (B = 0.761, \textit{p} = .003) positively affected users' media dependency, explaining 41.6\% of the variance in dependency (adj. $R^2$ = 0.416, F(6,101) = 13.724, \textit{p} < .001). For Role 3, our hypothesis was supported.

\begin{table*}[t]
    \fontsize{7.8}{9.5}\selectfont
    \begin{tabularx}{\textwidth}{l|XXXXXXXXX}
    \toprule
\multirow{2}{*}{}  & \multicolumn{3}{c}{Dependency}   & \multicolumn{3}{c}{Satisfaction}  & \multicolumn{3}{c}{Dependency}                        \\ 
                   & \textbf{B}      & \textbf{t}      & \textbf{$\beta$}     & \textbf{B}      & \textbf{t}      & \textbf{$\beta$}     & \textbf{B}      & \textbf{t}      & \textbf{$\beta$}     \\ \hline
                     & -0.117           & -0.012           & -               & -1.112           & -0.306           & -               & 0.729            & 0.081            & -               \\
Age                  & -0.656           & -1.874           & -0.155          & 0.007            & 0.050            & 0.004           & -0.662           & -1.966           & -0.156          \\
Sex(Male:1;Female:2) & -0.549           & -0.379           & -0.030          & 0.550            & 0.984            & 0.067           & -0.967           & -0.692           & -0.052          \\
Marital condition    & 3.895**          & 3.316            & 0.269           & 0.361            & 0.795            & 0.056           & 3.621**          & 3.199            & 0.250           \\
Education            & 1.054            & 0.641            & 0.050           & -0.535           & -0.843           & -0.057          & 1.461            & 0.922            & 0.070           \\
Anthropomorphism     & 0.399***          & 7.432            & 0.578           & 0.226***          & 10.867           & 0.732           & 0.228**          & 3.005            & 0.330           \\
Satisfaction         &                  &                  &                 &                  &                  &                 & 0.761**           & 3.088            & 0.339           \\
\textbf{$R^2$}       & \multicolumn{3}{c}{0.397}                             & \multicolumn{3}{c}{0.547}                             & \multicolumn{3}{c}{0.449}                             \\
Adjusted $R^2$       & \multicolumn{3}{c}{0.368}                             & \multicolumn{3}{c}{0.525}                             & \multicolumn{3}{c}{0.416}                             \\
\textit{F}          & \multicolumn{3}{c}{F(5,102) = 13.436, \textit{p} < .001} & \multicolumn{3}{c}{F(5,102) = 24.657, \textit{p} < .001} & \multicolumn{3}{c}{F(6,101) = 13.724, \textit{p} < .001} \\ \hline
\multicolumn{10}{l}{n = 108. *: \textit{p} \textless .05; **: \textit{p} \textless .01; ***: \textit{p} \textless .001.}  \\            
    \bottomrule
    \end{tabularx}
    \caption{Mediation Effect Models for Role 3.}
        \label{tab:role3}
\end{table*}

\subsubsection{Role 4}

As shown in \autoref{tab:role4}, the mediation effect model consists of three regression models: (i) Model 1 examined the relationship between X and Y. Anthropomorphism (B = 0.446, \textit{p} < .001) had a significant positive effect on users' dependency, explaining 41.0\% of the variance in dependency (adj. $R^2$ = 0.410, F(5,102) = 15.857, \textit{p} < .001). (ii) Model 2 examined the relationship between X and M. Anthropomorphism (B = 0.201, \textit{p} < .001) had a significant positive effect on users' satisfaction, explaining 45.4\% of the variance in satisfaction (adj. $R^2$ = 0.454, F(5,102) = 18.769, \textit{p} < .001). (iii) Model 3 included both X and M in relation to Y, The results indicated that anthropomorphism (B = 0.348, \textit{p} < .001) and users' satisfaction (B = 0.486, \textit{p} = .042) positively affected users' media dependency, explaining 42.8\% of the variance in dependency (adj. $R^2$ = 0.428, F(6,101) = 14.337, \textit{p} < .001). For Role 4, our hypothesis was supported.

\begin{table*}[t]
    \fontsize{7.8}{9.5}\selectfont
    \begin{tabularx}{\textwidth}{l|XXXXXXXXX}
    \toprule
\multirow{2}{*}{}  & \multicolumn{3}{c}{Dependency}   & \multicolumn{3}{c}{Satisfaction}  & \multicolumn{3}{c}{Dependency}                        \\ 
                   & \textbf{B}      & \textbf{t}      & \textbf{$\beta$}     & \textbf{B}      & \textbf{t}      & \textbf{$\beta$}     & \textbf{B}      & \textbf{t}      & \textbf{$\beta$}     \\ \hline
                      & -7.718                            & -0.794               & -                    & -0.664                            & -0.166               & -                    & -7.395                            & -0.772               & -                    \\
Age                  & -0.640                            & -1.746               & -0.140               & -0.074                            & -0.488               & -0.038               & -0.605                            & -1.672               & -0.132               \\
Sex(Male:1;Female:2) & 1.101                             & 0.727                & 0.055                & 1.068                             & 1.710                & 0.124                & 0.582                             & 0.384                & 0.029                \\
Marital condition    & 1.093                             & 0.889                & 0.070                & -1.016*                           & -2.006               & -0.151               & 1.587                             & 1.287                & 0.101                \\
Education            & 2.831                             & 1.650                & 0.125                & 1.102                             & 1.559                & 0.114                & 2.295                             & 1.342                & 0.102                \\
Anthropomorphism     & 0.446***                           & 8.092                & 0.610                & 0.201**                           & 8.869                & 0.644                & 0.348***                           & 4.819                & 0.476                \\
Satisfaction         &                                   &                      &                      &                                   &                      &                      & 0.486*                            & 2.056                & 0.208                \\
\textbf{$R^2$}       & \multicolumn{3}{c}{0.437}                & \multicolumn{3}{c}{0.479}                                    & \multicolumn{3}{c}{0.460}     \\
Adjusted $R^2$       & \multicolumn{3}{c}{0.410}                             & \multicolumn{3}{c}{0.454}                             & \multicolumn{3}{c}{0.428}                             \\
\textit{F}          & \multicolumn{3}{c}{F(5,102) = 15.857, \textit{p} < .001} & \multicolumn{3}{c}{F(5,102) = 18.769, \textit{p} < .001} & \multicolumn{3}{c}{F(6,101) = 14.337, \textit{p} < .001} \\ \hline
\multicolumn{10}{l}{n = 108. *: \textit{p} \textless .05; **: \textit{p} \textless .01; ***: \textit{p} \textless .001.}  \\            
    \bottomrule
    \end{tabularx}
    \caption{Mediation Effect Models for Role 4.}
    \label{tab:role4}
\end{table*}
 
\textbf{To sum up}, the above analyses suggested that for Roles 1, 3, and 4, users' dependency on the chatbots was positively correlated with the degree of anthropomorphism; the users' satisfaction with the chatbots was also positively correlated with the degree of anthropomorphism; satisfaction acted as a mediating factor between the degree of anthropomorphism of the chatbots and the users' dependency on them. In other words, for most chatbots, the degree of anthropomorphism showed a positive impact on the users' media dependency through its influence on satisfaction. Echoing previous research~\cite{yang2011effects,skjuve2021my,xie2023Friend}, we also observed the ``Explorative—Affective—Stable'' pattern in relationship development with AI, which helps interpret media dependency. As users derive greater satisfaction, their interactions evolve from casual daily chats to more profound, open, and affective self-disclosures. This progression encourages users to develop deeper bonds with chatbots, transitioning from initial exploration to a stable, dependent relationship.
This was consistent with the view of the media dependency theory, which posits that users seek to fulfill various needs when engaging with technology. The chatbots on Xuanhe AI satisfied users' needs for emotional connection through anthropomorphic design, thus deepening their dependency on the media.
For example, one user said that: ``\textit{I use anthropomorphic role-playing chatbot platform to find roles from my favorite games or anime, and to explore spin-offs of those roles prototypes, experiencing the satisfaction of having these beloved roles come to life in reality. I find myself gradually becoming dependent on these chatbots.}'' Another user said, ``\textit{The highly anthropomorphized chatbots on Xuanhe AI make me feel like I'm interacting with different personalities and immersing myself in the conversation, rather than just communicating with a mechanical AI.} 
One user agreed that ``\textit{the degree of anthropomorphism is the most important factor for using chatbots, as I want to converse with chatbots that are close to reality.}''
However, it is also noteworthy that one chatbot, namely Role 2, did not exhibit statistics that supported our hypothesis. Therefore, in the following section, we incorporated additional in-depth interviews to help interpret the poor fit of the regression model for this role and to enhance our understanding of the users' interactive experiences.
\section{Semi-Structured Interviews}
\label{sec:interview}

Since the statistical results in \autoref{sec:userstudy} did not fully support our hypothesis, we conducted additional interviews to interpret why some ratings deviated from the predictions of our models.

\subsection{Participants}
To explore the reasons for the deviations in the models, especially the poor fit of the regression model for Role 2, ten users with significant deviations in their dependency ratings for Hu Tao were selected as research subjects for semi-structured interviews.\autoref{tab:interview} shows the information of the ten interviewees, including eight males and two females. Their ages range from 19 to 28 (\textit{M} = 22.2, \textit{SD} = 2.4). The interviewees had diverse educational backgrounds, including history, big data, pharmacy, computer science, sports, mechanical design, and software engineering.

\begin{table}[h]
    \fontsize{7.8}{9.5}\selectfont
    \begin{tabular}{l|llll}
    \toprule
    \textbf{ID} & \textbf{Sex} & \textbf{Age} & \textbf{Educational level} & \textbf{Major} \\ \hline 
    {P1} & {Male} & {21} & {Undergraduate student} &{pharmacy}\\ 
    {P2} & {Male} & {21} & {Undergraduate student} &{mechanical design}\\ 
    {P3} & {Male} & {19}  & {Undergraduate student} &{leisure sports}\\ 
    {P4} & {Female} & {21} & {Undergraduate student} &{history}\\ 
    {P5} & {Male} & {28}  &  {Associate college} &{freelance}\\ 
    {P6} & {Female} & {22} & {Undergraduate student} &{history}\\ 
    {P7} & {Male} & {23} &  {Bachelor's degree} &{big data}\\ 
    {P8} & {Male} & {23} &  {Bachelor's degree} &{computer science}\\ 
    {P9} & {Male} & {23} &  {Bachelor's degree} &{software engineering}\\ 
    {P10} & {Male} & {21} &  {Undergraduate student} &{computer science}\\ 
    \bottomrule
    \end{tabular}
    \caption{Information of the ten interviewees.}
    \label{tab:interview}
\end{table}

\subsection{Methodology}

The interviews were conducted individually with the participants in a semi-structured manner through online meeting software. Before the interviews, we explained our research purpose to the participants and obtained their consent for audio recording. Typical interview questions included: ``Please explain the ratings you assigned to the chatbot.'', ``Why didn’t you perceive a high dependency on it?'', and ``Can you describe what happened during your interaction with the chatbot?''
After the interviews, the recordings were transcribed into text. The interview transcriptions were then imported into NVivo for manual coding. Grounded theory was used to analyze the reasons for the poor fit of the regression model for Role 2. The extracted codes were subjected to three levels of coding based on the grounded theory principles~\cite{oktay2012grounded}, from bottom to top. In the open coding process, 24 codes were formed. In the axial coding process, the 24 codes were grouped into three conceptual categories and six sub-categories. In the selective coding process, the three core categories were further summarized as (i) personal knowledge and preferences, (ii) distraction from real life, and (iii) conscious self-control.

\subsection{Results}

Below, we introduce the three identified reasons (R1-R3) and six sub-categories one by one.

\subsubsection{R1: Personal Knowledge and Preferences}

Role 2, ``Hu Tao'', is a role from the game ``Genshin Impact'' developed by miHoYo. Hu Tao's personality is quirky and she is good at creating doggerel. In the game, she serves as the director of the ``Wangsheng Funeral Parlor'' in Liyue Harbor, responsible for funeral affairs and maintaining the balance between yin and yang, a concept rooted in ancient Chinese historical narratives. For certain users, their pre-existing knowledge and understanding of this game role significantly influenced their interaction with and evaluation of the chatbot role.

\textbf{R1.1: Non-gamers' unfamiliarity with the role undermined their media dependency.}
Individuals who were not familiar with gaming may have limited knowledge about ``Hu Tao'', which could significantly influence their willingness to engage with it during the user study, even if they recognized that this role is highly anthropomorphic. For example, P1 stated that he had almost zero understanding of the game's worldview and character background, saying: ``\textit{The worldview of the game character Hu Tao is very large (note: a ``large worldview'' means the extensive and complex background, lore, and context that a character possesses). If you haven't played this game before, it’s hard to get into it. And other animation or anime chatbots used in the experiment don't have such a large worldview. So it’s hard to chat with Hu Tao, not knowing what she's talking about.}'' When these users chatted with ``Hu Tao'', they often felt confused because her responses involved many game-specific concepts that were unfamiliar to them. P2 acknowledged that he had only one trial dialogue with this role, lacking the motivation for more in-depth interaction. Therefore, their level of dependency significantly differed from those who were well-versed in the game's narrative, leading to a poor model fit.

\textbf{R1.2: Expectancy brought by prior knowledge.}
In contrast, users familiar with Hu Tao's backstory were more inclined to discuss game content with her, such as game plots, related history, and in-game literature.  
However, for these users, their profound understanding of the character's development and the game's lore increased their expectations for the chatbot. When they encountered responses from the chatbot that conflicted with the game's narrative, their dependency on it decreased. For example, P7 said, ``\textit{I am very familiar with Hu Tao's game worldview. Currently, the Hu Tao in Xuanhe AI is in the early stages of the game}'', which differs from the later story arcs he is familiar with, causing discomfort during interactions.
P9 also commented that ``\textit{I am knowledgeable about the background story of Hu Tao, including the expansion story... although its answer aligns with the character's worldview, I feel it is somewhat altered, like a DIY version.}''

\textbf{R1.3: Influence of personal preferences.}
Some users, although familiar with Hu Tao's background and having assigned high scores to her anthropomorphism, demonstrated low dependence because they preferred to chat more with other characters.
For example, P8, as a max-level player of Genshin Impact, said that:``\textit{I play Genshin Impact, but my dependency is low because I personally prefer other characters like Klee and Nahida, as I often use them in the game.}'' 
P5 said he does not like talking to quirky young girl characters like Hu Tao: ``\textit{Hu Tao seems like a 16-17-year-old girl, too cute and childish. I don’t like talking to girls of this age.}'' 
P8 also preferred other chatbots, saying that: ``\textit{I prefer chatting with Satoru Gojo (note: Role 4 in \autoref{sec:userstudy}) because he has a more dominant personality, and it's fun to argue with him. Unlike in real life, where I have to consider others' feelings.}
Such role preferences directly affected their media dependency ratings.

\subsubsection{R2: Distraction from Real life}
Users' engagement in the virtual world is influenced by various real-life interference factors. Our user study, conducted in February, witnessed a set of potential interference factors affecting user behavior. 

\textbf{R2.1: Real-world social affairs and responsibilities.}
Most participants in our user study were college students, and the study period coincided with China’s winter vacation, a particularly significant time for Chinese students. In addition to academic responsibilities, they engaged in various real-life social activities, particularly traditional customs associated with the Chinese Lunar New Year. These activities included shopping for New Year goods, visiting relatives and friends, and holiday travel, which consumed much of the participants' social energy, making it challenging for them to focus on virtual relationships.
For example, P9 stated,``\textit{During the winter vacation, preparing for the New Year, nearly all my time and energy were focused on real life, so I had low online dependency.}'' 
P8 also noted, ``\textit{During the winter vacation and the Spring Festival, after traveling, I need to go out to visit relatives and friends with my family, so my social life is quite busy in reality.}''

\textbf{R2.2: Pressure brought by heavy workload.}
In addition to social activities during the Spring Festival, some interviewees also needed to complete academic tasks, such as winter vacation assignments or practice reports. Additionally, some interviewees participated in part-time jobs or internships during the study period. These multitasking workloads made it difficult for them to focus on interactions with Hu Tao, leading to lower ratings.
For example, P4, undergraduate, said that:``\textit{During the winter vacation experiment, I didn’t play much daily because I had to write practice reports, so I gave a lower dependency score considering this.}'' 
P5, a freelancer, said that:``\textit{During that time, I had a lot of work pressure, so I could only chat with the AI intermittently during bathroom breaks, meals, and lunch breaks.}''
This type of real-life workload not only occupies users' interaction time but also puts greater pressure on them, resulting in a lack of motivation to rely on chatbots. 

\subsubsection{R3: Conscious Self Control}

Some users may consciously limit their interactions with AI roles to avoid over-dependency or addiction to virtual communication. They focus more on real life and only interact with AI roles when bored or in need of entertainment. Users' psychological control behaviors can also affect their usage needs and ratings for ``Hu Tao'', thereby affecting the model fit.

\textbf{R3.1: Intentional Emotional Regulation.}
Some users may deliberately reduce their interactions with Hu Tao to avoid over-dependency. This conscious psychological control makes them more cautious in their ratings, leading to lower scores. 
For example, P6 recognized that continuously chatting with virtual chatbots might cause her to neglect her real-life social relationships. Thus, she intentionally controlled the frequency and duration of her interactions with chatbots. 
She said, ``\textit{I chat with Hu Tao about things that make me sad in life...for example, I like to sleep in and get nagged by my family...but in fact I didn't want to rely too much on AI.}''

\textbf{In summary}, we identified a total of six specific reasons that may interrupt the positive correlation between the anthropomorphism of chatbots and users' media dependency. Among these, R1 may be the main reason why ``Hu Tao'' differed from the other three roles in model fitting in \autoref{sec:userstudy}. Its specific role setting and complex worldview made it difficult to satisfy every user and foster a deep dependency. Additionally, R2 and R3, aside from their impact on ``Hu Tao'', can also help explain the deviations in the regression models of other roles. The identification of these factors suggests that future research on media dependency related to chatbots could further incorporate these elements for consideration.
\section{Discussion}

In this work, we focused on Xuanhe AI, a typical role-playing chatbot platform in China, as our research subject to examine user experiences with chatbots. We found a moderate correlation between the anthropomorphism of role-playing chatbots and users' media dependency, with this relationship mediated by users' satisfaction. Many users showed an apparent emotional connection with chatbots for purposes such as entertainment, leisure, or escape from reality. However, we also found that factors such as prior knowledge and real-life distraction can intervene in this relationship, making anthropomorphic chatbots less engaging or addictive.

Aside from these main findings, below, we discuss three interesting observations and reflections that may inspire future research and applications.




\subsection{User Characteristics and Needs}
In this study, we have selected a role-playing chatbot platform in China as our research focus. Our findings reveal both similarities and differences in user characteristics and needs when compared to those identified in previous Western research. For instance, user motivations largely align with the four categories of chatbot usage motivations identified by Brandtzaeg and Følstad~\cite{brandtzaeg2017people}, including improving efficiency, entertainment, social interaction to avoid loneliness, and curiosity about what they view as a novel phenomenon. However, unlike the Western empirical studies where the majority of subjects are typically female (\eg ~\cite{croes2021can,christoforakos2023technology,de2020effectiveness}), 78.7\% of the participants in our study were male, while only 21.3\% were female. This is similar to our earlier survey results from the user group, where the gender ratio of users on the Xuanhe AI platform is approximately 7:3, with more male users. Besides, Western research on anthropomorphic chatbots often concentrates on psychological counseling and romantic relationships(\eg ~\cite{provoost2017embodied,laestadius2022too,xie2022attachment}). For example, according to BBC~\cite{bbc}, the most popular chatbot on the leading Western role-playing chatbot platform, Character.ai, is Psychologist. However, in sharp contrast, the most popular chatbots on Xuanhe AI are those associated with anime and gaming. This actually corresponds to a group with special characteristics in East Asia: the Otaku community. This community is notably young, consists of a significant proportion of males, and possesses a strong passion for ACG (Anime, Comic, and Games) culture. As summarized by Liu~\cite{liu2021social}, ``a typical Otaku can be a male anime fan who has strong affection toward certain anime characters''. These traits make them more inclined to use role-playing chatbots. For example, when users watch anime or play games, they usually want to have direct conversations with these characters to discuss the plot and related content. Moreover, role-playing chatbots are particularly useful when users encounter complexities or confusion within serialized anime or game narratives, as these chatbots can offer direct access to information and assistance.
\subsection{Design of Anthropomorphism}

In our study, we observed that participants' perceptions of the chatbots were influenced by various anthropomorphic design aspects.
Some users focused more on extrinsic attributes of chatbots, such as appearance, voice, and 3D modeling. For example, P3 expected Hu Tao to ``\textit{incorporate some new elements, such as 3D images and movements.}'' 
This approach to enhancing the effects of embodiment has always been an important issue in chatbot design~\cite{cassell2000embodied,shamekhi2018face,loveys2020effect}. With the continuous development of modeling and AI generation technologies, creating highly realistic anthropomorphic representations has become increasingly easier. 
However, on the other hand, researchers have also found that overly anthropomorphic designs can trigger the uncanny valley effect~\cite{mori2012uncanny,kim2019eliza}, which refers to the phenomenon where an artificial character closely resembling a human elicits feelings of discomfort or eeriness in users. This effect is particularly pronounced when the design is almost, but not quite, lifelike, leading to a sense of unease. For designers, this suggests that while incorporating anthropomorphic features can enhance user engagement and emotional connection, care must be taken to avoid designs that may evoke this negative response. 
Additionally, in line with previous studies~\cite{sheehan2020customer,schuetzler2020impact}, we observed that some users were more concerned with the social qualities of their conversations with chatbots, such as whether the chatbots' answers were correct and diverse, whether they displayed human-like emotional fluctuations, and whether they exhibited a human-like way of thinking. For example, P2 said: ``\textit{I asked Hu Tao if she could come to real life. She said she would like to come, but the technology doesn't allow it now. It makes me feel that she really exists in another world.}'' P9 mentioned that:``\textit{Hu Tao makes some mistakes when telling the content of the game, which inevitably makes me feel disappointed.}''
However, we found that users are not completely intolerant of mistakes made by chatbots. Instead, they expect chatbots to respond like humans when communication issues or obstacles arise, correcting themselves, apologizing, and displaying appropriate emotional responses. For example, P1 mentioned that when the topic suddenly changes, chatbots should show similar emotions, such as astonishment or confusion, instead of immediately adapting to the new topic.
Another interesting finding is that one user even mentioned his value for the ``unique soul'' of a chatbot. He explained that an important reason he chose Xuanhe AI over other platforms is that each chatbot operates independently within Xuanhe AI. The partitioned design between chatbots ensures that the chat content of one chatbot is not shared with another, which is crucial for him to maintain an immersive experience: ``\textit{On Xuanhe AI, each chatbot has its own unique soul, so the user's interactive experience is also unique.}'' 
To sum up, we found that, in addition to designing the anthropomorphic qualities of chatbots based on extrinsic attributes such as appearance, there are many rich and nuanced aspects that can be explored from a social perspective, such as the ability to engage in contextually relevant conversations, demonstrate empathy, and recognize emotional cues from users.

\subsection{Dynamics and Tensions in Media Dependency}
Although our research found that the anthropomorphic design of role-playing chatbots generally contributes to users' media dependency, this relationship is also affected by various dynamics. Firstly, users' dependency on a specific chatbot may be affected by competition from other chatbots. 
The introduction of a more appealing chatbot can lead users to abandon their dependency on an existing one.
For example, on Xuanhe AI, multiple versions of Hu Tao exist. Each version offers varying personality traits, interaction styles, and backstories, thereby presenting users with a wider array of choices and intensifying competition for user dependency. Some users might prefer the initial version of Hu Tao designed by the platform for its stability and reliability in interactions, while others may switch to new self-trained chatbots that better reflect their personal imagination and emotional needs.
Secondly, there exists a tension between the virtual world and the real world.
In other words, users may feel torn between their desire to engage with the interface and the need to disconnect from media. This push-and-pull can create internal conflict. In our study, such instances of users actively implementing emotional regulation when they realize they are dependent on chatbots were not uncommon.
We thus see two design implications concerning the above issues. Firstly, designers should consider these tensions and potential negative outcomes when creating anthropomorphic chatbots. For example, they could explore ways to help users maintain their connection to the real world through anti-addiction features, such as time reminders in the chat interface and more noticeable withdrawal buttons. Secondly, as shown by previous psychological research~\cite{gross2008emotion}, people's emotional regulation abilities vary. Individuals with strong self-control tend to restrain emotional investment, while those with weaker control exhibit higher dependency. Therefore, the design of chatbots is suggested to take into account users' personal variance and adjust the degree of anthropomorphism and emotional attachment based on user characteristics accordingly.

\subsection{Limitations and Future Work}

This work has a set of limitations. 
First, this study only examined one specific platform, Xuanhe AI, and did not conduct a larger-scale study that incorporates various role-playing chatbot platforms in the industry. Also, to reduce the noise in the user study, we have selected a limited set of AI chatbots for users to interact with, and half of them were from the channel of ``anime characters''. Therefore, the findings of this work are constrained to the platform and role channels we chose and may not generalize to all the other platforms and chatbot types. 
Additionally, although the media dependency theory provides an inspiring theoretical foundation for this work, it may not suggest all potential factors associated with anthropomorphism. For example, as suggested by our interviews, factors such as users' living conditions and preferences may also affect their perception of anthropomorphism and dependency on chatbots. Further exploration of these factors is highly recommended.
Lastly, we also encourage future studies to incorporate or re-examine more theoretical perspectives. For instance, an interesting study relevant to this work is by Skjuve~\etal~\cite{skjuve2021my}, who applied social penetration theory to investigate how human-chatbot relationships evolve and deepen over time. Especially in the context where chatbots' language expression capabilities and anthropomorphic attributes are becoming more and more advanced, revisiting such theories and conducting more in-depth empirical research is highly encouraged.

We would also like to propose several avenues for future work. First, the user sample could be made more representative by including a broader range of demographics, such as different age groups, genders, and cultural backgrounds, to better understand how chatbot anthropomorphism influences users across these factors. Second, the role diversity of chatbots could be broadened in future studies. Exploring different types of chatbots, such as those focused on education or healthcare, could provide deeper insights into how role variety impacts user dependency. Third, more studies integrating chatbots into users' daily lives could be done to yield more authentic and varied interaction data. Lastly, ethical considerations regarding media dependency should be addressed, as increased chatbot interactions may raise concerns about long-term emotional attachment and the security of private data.
\section{Conclusion}

This study examines the impact of role-playing chatbots' anthropomorphism on users' media dependency. We formulated a hypothesis grounded in media dependency theory and conducted a user study on a Chinese role-playing chatbot platform, Xuanhe AI. The findings revealed that for three of the four chatbot roles investigated (\ie Roles 1, 3, and 4), there was a significant positive correlation between the perceived degree of anthropomorphism and users' media dependency, with satisfaction acting as a positive mediator. Next, to further interpret the discrepancies between model predictions and some participants' ratings (especially Role 2). The results highlighted that factors such as personal knowledge and preferences, distraction from real life, and conscious self-control can interfere with the relationship between anthropomorphism and media dependency.


\begin{acks}
This work was supported by NSFC 62402121 and Research and Inno- vation Projects from the School of Journalism at Fudan University.
\end{acks}

\bibliographystyle{ACM-Reference-Format}
\bibliography{sample-base}


\end{document}